\documentclass[12pt]{iopart}
\usepackage{amssymb}
\usepackage{graphicx}
\usepackage{bm}
\usepackage{hyperref}
\newcommand{\ot}{\tilde{\Omega}}

\begin{document}
\title{Kinetic analysis of a chiral granular motor}
\author{Julian Talbot, Alexis Burdeau and Pascal Viot} 
\address{Laboratoire de Physique
  Théorique de la Matière Condens\'ee, Universit\'e Pierre et
  Marie Curie, CNRS UMR 7600, 4, place Jussieu, 75252 Paris Cedex 05, France}
\ead{talbot@lptmc.jussieu.fr, viot@lptmc.jussieu.fr}

\bibliographystyle{unsrt}

\begin{abstract} 
We study the properties of a heterogeneous, chiral granular rotor that is capable of
performing useful work when immersed in a bath of thermalized particles. 
The dynamics  can be obtained
in general from a numerical solution of the  
Boltzmann-Lorentz equation. We show that a
mechanical approach gives the exact mean angular velocity in the limit
of an infinitely massive rotor.  We examine the dependence of the mean angular
velocity on the coefficients of restitution of the two materials composing
the motor. We compute the power and efficiency and compare with 
numerical simulations. We also perform a realistic numerical simulation of a
granular rotor which shows that the presence of non uniformity of the bath
density within the region where the motor rotates, and that the ratchet effect
is slightly weakened, but qualitatively sustained. Finally we discuss the results
in connection with recent experiments. 
\end{abstract}
\maketitle

\section{Introduction}

Much is known about the properties of Brownian ratchets and
their applications in physics and biology.  
Smoluchowski's \cite{smolu12} original proposal consisted of  
a device with four vanes
connected to a ratchet and pawl. When immersed in a thermalized 
bath of particles the device can apparently rectify the thermal fluctuations
to produce useful work. Feynmann\cite{feymann63}, however, demonstrated that
this is not possible if the entire system is at thermal equilibrium. 
The device only performs work if the vanes and the ratchet and pawl are
maintained at different temperatures, $T_1$ and $T_2$, respectively with 
$T_1>T_2$.  
It is now understood that the general requirements for a functioning motor  
are the absence of 
both time and spatial symmetry. 
The former occurs if there is
a breakdown of detailed balance and the latter may be due to 
the intrinsic asymmetry of the object itself or to an external
force \cite{Reimann2002,broek:011102,Broek2009}.

In the last few years, several proposals for granular motors
have appeared. The dissipative nature of inelastic collisions
leads to an automatic breakdown of time reversal symmetry and the spatial
asymmetry can be created in various ways. 
In 2007 Cleuren and Van den Broeck \cite{Cleuren2007} and independently, Costantini, 
Marconi and Puglisi \cite{Costantini2008,Costantini2009} proposed the same
model granular motor, viz.  
an isosceles triangle composed of a homogeneous
inelastic material constrained to
move along a line. When immersed in a bath of
thermalized particles, the undirected fluctuations
of the bath particles induce a directed
motion of the triangle. 
The drift velocity is proportional to $(1-\alpha)\sqrt{m/M}\sqrt{\frac{kT}{M}}$ where $\alpha$ is the coefficient of 
restitution characterizing the collision between the tagged particle (of mass $M$) and the bath particles (of mass $m$).
This result shows that the phenomenon is specific to granular particles, because the drift velocity vanishes when 
$\alpha=1$. Also, because the granular temperatures of the tagged particle and the bath particles
are comparable, whatever the mass ratio, $<(V-<V>)^2>$ decreases as $m/M$ (for a given bath temperature), 
and this results in a signal-to-noise ratio that vanishes for large values of $\sqrt{m/M}$.

Cleuren and Eichhorn \cite{Cleuren2008} later proposed an alternative model consisting
of a homogeneous rotor that is constrained to rotate about an off-center axis.
If the material composing the rotor is inelastic, a net rotation is obtained and it displays the same dependence on $\alpha$ 
and $m/M$ as the triangle, which complicates the experimental observation of this phenomena.

Fortunately, this difficulty can be overcome by using
a heterogeneous device constructed from two different
materials with different coefficients of restitution. This
leads to a strong ratchet effect \cite{granada2010}.
Costantini {\it et al.} \cite{costantini:061124} proposed possibly the
simplest model of a granular motor of this type, the asymmetric piston
composed of two materials with different inelasticities.  Starting from 
the Boltzmann equation, they proposed a phenomenological
approach based on the evolution of the first three
moments of the velocity distribution. Although their theory is 
in good agreement with numerical simulation results in some cases, 
it breaks down in the limit of large piston mass. 

Recently, Eshuis {\it et al}. \cite{PhysRevLett.104.248001} succeeded in constructing a macroscopic rotational
ratchet consisting of four vanes that rotates in a granular gas.  They performed two kinds of experiments: 
in the first, the vanes were symmetric and no net angular velocity was observed. In the second, 
asymmetry was induced by coating each side of the vanes 
differently. For sufficiently large granular temperatures of the bath particles, a net rotation is observed.  
We will discuss the implications of this experiment in the concluding section.

In a recent article \cite{PhysRevE.82.011135}  we presented a mechanical
approach that gives the average force acting on a heterogeneous
particle moving at a given velocity. The mechanical approach is consistent with 
the Boltzmann equation and in the Brownian limit, 
$m/M\rightarrow 0$, it leads directly to an {\it exact}
expression for steady state drift velocity.   

From both technological and fundamental perspectives, the power and
efficiency of thermal engines are of major interest.  Indeed, Carnot's
analysis of an idealized heat engine is a landmark
event in the history of thermodynamics. His famous result for the efficiency depends only
the temperatures of the heat reservoirs, but applies strictly to a
reversible - and hence quasi-static - process. The efficiency at
non-zero power was first considered by Curzon and
Ahlborn\cite{curzon:22} and later generalized by van den Broeck
\cite{PhysRevLett.95.190602}.  The efficiency of Brownian motors has
also been discussed \cite{RevModPhys.69.1269,Schmiedl2008}, but there has
been little discussion of the 
power and efficiency of granular motors \cite{PhysRevE.82.011135}. Such 
considerations are, nevertheless, of great importance for 
the practical realization of these devices.

It is the purpose of this article to apply the approach first proposed in 
\cite{PhysRevE.82.011135} to the chiral rotor. We also investigate the effect
of bath inhomogeneities with a realistic numerical
experiment. Finally, we show that the mechanical approach provides an
explanation for the presence of a threshold for the rotation of the vanes in the
experiment of Eshuis et al. \cite{PhysRevLett.104.248001}.

\begin{figure}[t]
\begin{center}
\resizebox{12cm}{!}{\includegraphics{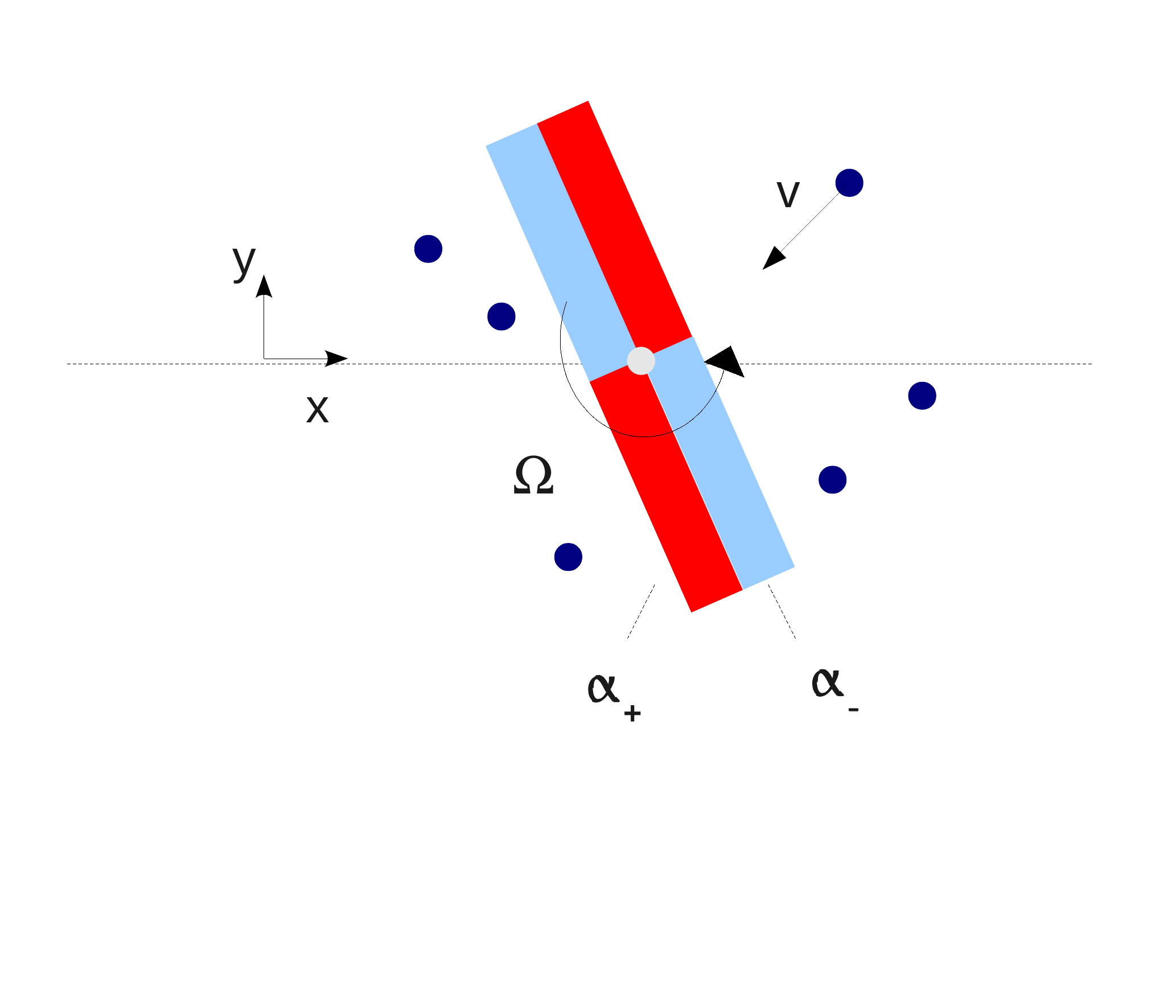}}
\end{center}
\caption{
  The chiral rotor is constructed from two different materials with
  coefficients of restitution $\alpha_{+}$ (red) and $\alpha_{-}$ (blue). When 
immersed in a bath of thermalized particles, it acquires a net
  rotation around its axis.}
\label{fig:rotor}
\end{figure}

\section{Model}

The chiral rotor is composed of two materials with 
coefficients of restitution $\alpha_{+}$ and $\alpha_{-}$: See figure~\ref{fig:rotor}. Collisions of
the bath particles with the former result in a positive torque (anticlockwise sense), while collisions with the material
with $\alpha_{-}$ result in a negative torque.  The device is immersed in a two-dimensional granular gas composed 
of structureless particles at a density $\rho$ each of mass $m$. 

An important assumption of the model is that
collisions between the motor and the bath particles do not modify the velocity distribution of the latter. This
neglect of recollisions is consistent with the foundations of Boltzmann's kinetic theory. We let $v_\perp$ and
$v_\parallel$ denote the components of the gas particle's velocity
perpendicular and parallel to the surface of the motor, respectively. The granular
temperature of the bath is defined as $T_B=m<v_\perp^2>=m<v_\parallel^2>$.    
Since $v_\parallel$ is irrelevant in the Boltzmann-Lorentz description of the rotor\cite{piasecki:051307}, for sake of
simplicity we denote $v_\perp$ by $v$ in the rest of the section.

Wherever possible, we will give results for an arbitrary bath particle velocity
distribution, $\phi(v)=\int dv_\parallel \phi({\bf v})$. Sometimes, though, it will be necessary to assume a particular form and in this case we will use a  
Maxwell-Boltzmann or
Gaussian distribution:
\begin{equation}\label{eq:MB}
\phi_{MB}(v)=\frac{1}{\sqrt{2\pi}v_{th}}\exp\left(-\frac{v^2}{2v_{th}^2}\right),
 \end{equation}
where $v_{th}=\sqrt{T_B/m}$. 
While granular gases often have non-Maxwellian velocity distributions,
it is possible to devise experimental situations where the Gaussian
distribution is observed \cite{Baxter2003,baxter:028001}.

The collision equations are
\begin{eqnarray}
\left( \begin{array}{c}
\Omega' \\
v' \\
\end{array} \right)=\left( \begin{array}{c}
\Omega \\
v \\
\end{array} \right)+\frac{1+\alpha}{I+mx^2}(v-\Omega x)
\left( \begin{array}{c}
mx \\
-I \\
\end{array} \right)
\end{eqnarray}
where $\Omega$ is the angular velocity, $I$ is the moment of inertia and $-L/2 \le x\le L/2$ is the algebraic 
distance of impact  from the center (for simplicity, we neglect collisions between bath particles and the caps of the rotor). The
granular temperature of the rotor is given by $T_g = I<(\Omega-<\Omega>)^2>$. Because the mass of any section of thickness $\Delta x$
perpendicular to the long axis of the rotor is constant, the moment of inertia
is the same as for a homogeneous rotor, $I=ML^2/12$, where $M$ is the total mass. 
The energy loss on collision is
\begin{equation}\label{eq:energy} 
 \Delta E=-\frac{mI(1-\alpha^2)(v-\Omega x)^2}{2(I+mx^2)}
\end{equation}

The reconstituting velocities are 
\begin{eqnarray}
\left( \begin{array}{c}
\Omega^{**} \\
v^{**} \\
\end{array} \right)=\left( \begin{array}{c}
\Omega \\
v \\
\end{array} \right)+\frac{1+\alpha^{-1}}{I+mx^2}(v-\Omega x)
\left( \begin{array}{c}
mx \\
-I \\
\end{array} \right)
\end{eqnarray}

\section{Boltzmann-Lorentz Equation}
The time dependent Boltzmann-Lorentz equation for the chiral rotor is
\begin{eqnarray}\label{eq:berot0}
\fl\frac{\partial}{\partial t} f(\Omega;t) &=\rho
\int_{-L/2}^{L/2}dx\,\int_{-\infty}^{\infty}dv\;
|v-x\Omega|\left[\theta(v-x\Omega)\frac{f(\Omega^{**};t)}{\alpha_{+}^2}\phi(v^{**})
+\theta(x\Omega-v)\frac{f(\Omega^{**};t)}{\alpha_{-}^2}\phi(v^{**})\right]\nonumber\\
&-\rho f(\Omega;t)\int_{-L/2}^{L/2}dx\,\int_{-\infty}^{\infty}dv\;|v-x\Omega|\phi(v)
\end{eqnarray}
where $\theta(x)$ is the Heaviside function. 
By introducing the variable $y=(v-\Omega x)/ \alpha_{+}$, $y=(\Omega x-v)/ \alpha_{-}$ and 
exploiting the symmetry of the object 
we can write the equation in the more explicit and useful form:
\begin{eqnarray}\label{eq:berot}
\fl \frac{1}{2\rho}\frac{\partial}{\partial t} f(\Omega;t) &=
\int_{0}^{L/2}dx\,\int_0^{\infty}dy\;
yf\left(\Omega-\frac{1+\alpha_{+}}{I+mx^2}mxy\right)\phi\left(\Omega
x+\frac{I-\alpha_{+}mx^2}{I+mx^2}y\right)\nonumber \\
 &+\int_{0}^{L/2}dx\,\int_0^{\infty}dy\;yf\left(\Omega+\frac{1+\alpha_{-}}{I+mx^2
} mxy\right)\phi\left(\Omega x-\frac{I-\alpha_{-}mx^2}{I+mx^2}y\right)\nonumber
\\
&-f(\Omega)\int_{0}^{L/2}dx\,\int_0^{\infty}dy\;y(\phi(\Omega x+y)+\phi(\Omega x-y))
\end{eqnarray}

When the bath distribution is Gaussian, equation \ref{eq:MB}, 
one can confirm that for a rotor with $\alpha_{+}=\alpha_{-}=1$, the solution in the steady state is given by $f(\Omega)=\sqrt{I/2\pi T_{B}}\exp(-I\Omega^2/2T_{B})$.

\subsection{Numerical solution}
The Direct Simulation Monte Carlo (DSMC) method \cite{B94}  is often used to obtain numerical solutions of the Boltzmann equation 
and has been applied to granular systems \cite{MS00}. An alternative is the Gillespie method \cite{TV06}. 
In this application, the linear nature of the Boltzmann-Lorentz equation makes it well suited to solution
by iteration. To achieve this, we rewrite the equation in the following form: 
\begin{eqnarray}
\fl f^{(n+1)}(\Omega)&=\frac{1}{\nu(\Omega)}
\left[\int_{0}^{L/2}dx\,\int_0^{\infty}dy\;
yf^{(n)}\left(\Omega-\frac{1+\alpha_{+}}{I+mx^2}mxy\right)\phi\left(\Omega
x+\frac{I-\alpha_{+}mx^2}{I+mx^2}y\right)\right.\nonumber \\
&\left.+\int_{0}^{L/2}dx\,\int_0^{\infty}dy\;yf^{(n)}\left(\Omega+\frac{1+\alpha_{-}}{I+mx^2
} mxy\right)\phi\left(\Omega x-\frac{I-\alpha_{-}mx^2}{I+mx^2}y\right)\right]
\end{eqnarray}
where $f^{(n)}$ denotes the $n$th iteration and
\begin{equation}
\nu(\Omega)=\int_{0}^{L/2}dx\,\int_0^{\infty}dy\;y(\phi(\Omega x+y)+\phi(\Omega x-y))
\end{equation}
so that $2\rho\nu(\Omega)$ is the collision rate when the motor rotates with angular velocity $\Omega$. 
We take as the initial guess, $f^{(0)}$, either a Gaussian distribution or
the previous converged solution for different parameters. The error is defined as
\begin{equation}
\epsilon=\int_{-\infty}^{\infty}[f^{(n+1)}(\Omega)-f^{(n)}(\Omega)]^2\,d\Omega
\end{equation}
In the results reported here we took the solution as converged when $\epsilon<10^{-15}$.

\begin{figure}[ht]
\begin{center}
\resizebox{8cm}{!}{\includegraphics{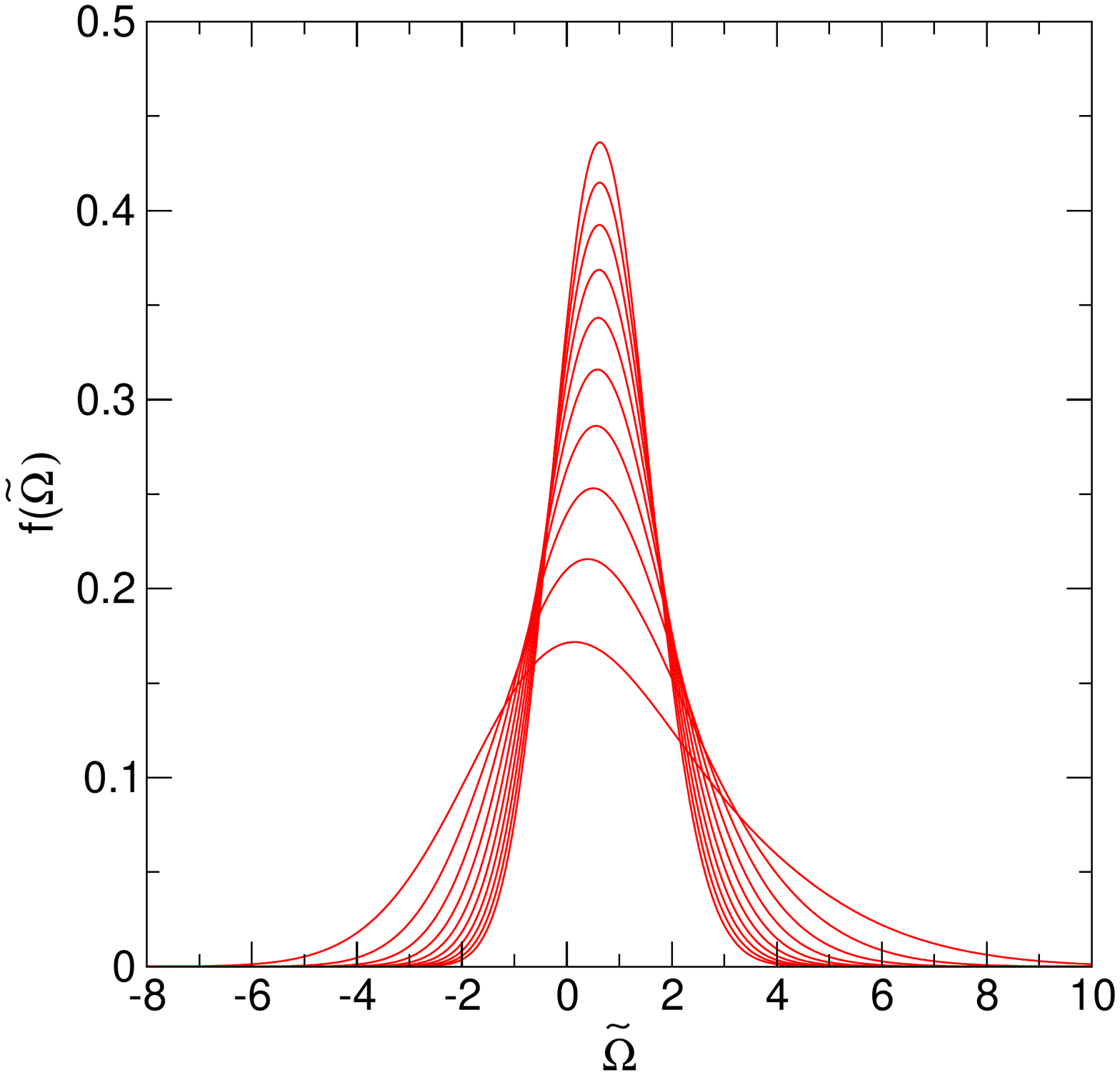}}\\
\resizebox{8cm}{!}{\includegraphics{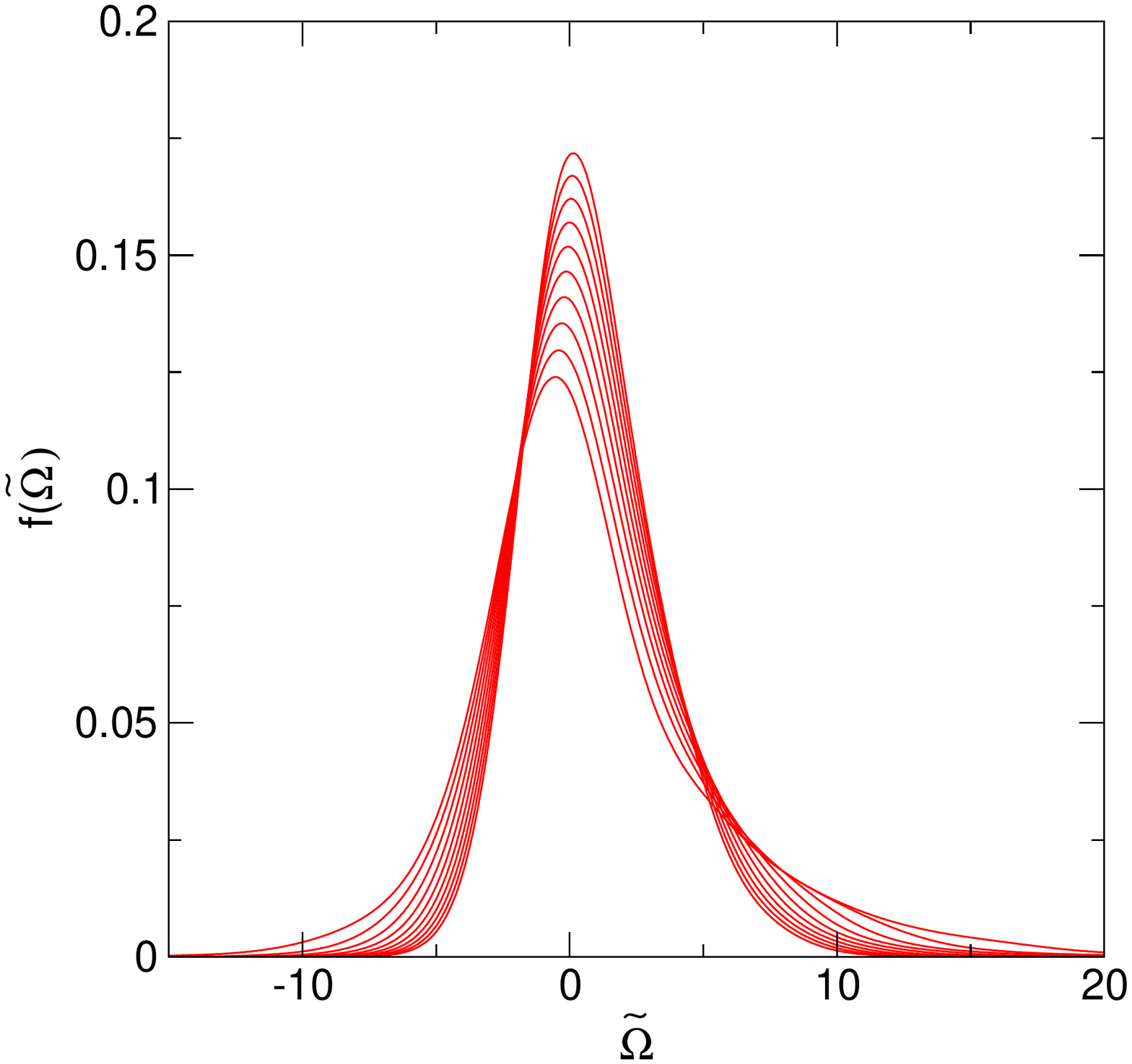}}
\end{center}
\caption{Angular velocity distributions 
for a chiral rotor with $\alpha_{-}=0$, $\alpha_{+}=1$ obtained by iteration of the Boltzmann equation. 
Top: $1\le I \le 10$ in steps of 1; Bottom $0.1\le I \le 1.0$ in steps of 0.1}
\label{fig:ber}
\end{figure}

We show some results in figure \ref{fig:ber}.  
As the rotor becomes heavier, the distribution becomes more Gaussian with the maximum at the drift velocity $\Omega^*$. 
Conversely a light rotor has a markedly non-Gaussian distribution that becomes more and more skewed with decreasing
mass. For a sufficiently light rotor, the most probable angular velocity is opposite in
sign to the mean value. This is similar to the behavior observed for the asymmetric piston \cite{Costantini2008}.

\begin{figure}[ht]
\begin{center}
\resizebox{8.0cm}{!}{\includegraphics{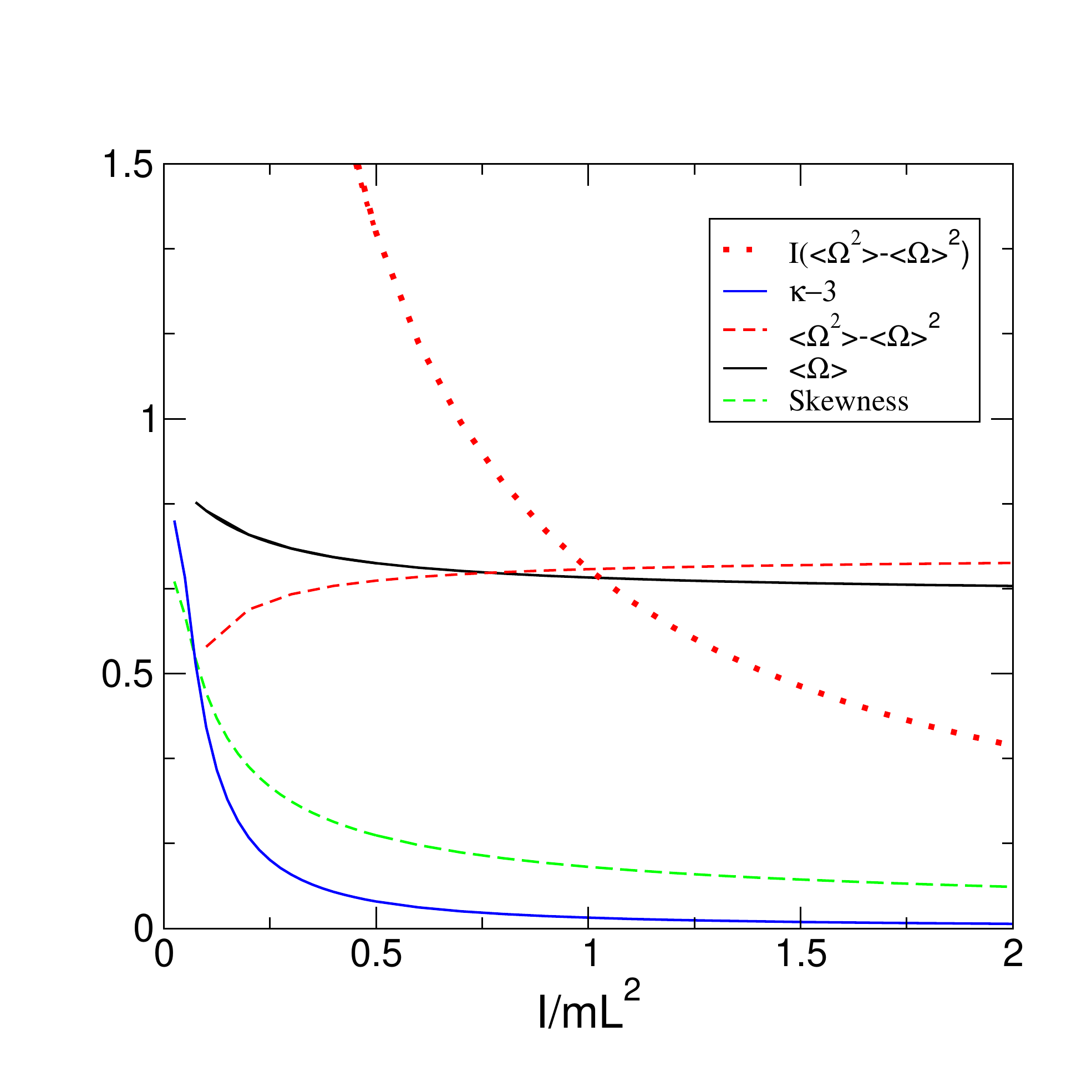}}
\end{center}
\caption{First moment (second from top, right hand side), second moment (third from bottom, right hand side), skewness (second from bottom, right hand side) and excess kurtosis (bottom curve) of the velocity distribution for $\alpha_{-}=0$, $\alpha_{+}=1$ as a function of $\mu$. 
The dashed line shows the granular temperature of the rotor.}.
\label{fig:Rmoments}
\end{figure}

The increasingly non-Gaussian behavior can be quantified by examining the skewness and kurtosis, $\kappa$, 
 of the distribution: See
figure \ref{fig:Rmoments} (for convenience the excess kurtois, defined as $\kappa-3$, is shown). While the excess 
kurtosis is nearly zero beyond $I/mL^2=1$, there is some
residual skewness for large rotor masses. We also note the the granular temperature of the rotor varies little beyond  $I/mL^2=1$.

\subsection{Kramers-Moyal expansion}

Henceforth we focus on the Brownian limit as it is more interesting for the experimental application. 
As with the piston\cite{PhysRevE.82.011135} we can perform a series expansion of 
Boltzmann equation by introducing  the small parameter
$\epsilon=\sqrt{\frac{mL^2}{I}}$, the variable $z=(\Omega-\Omega^*)/ \epsilon$ 
where $\Omega^*$ is  the drift velocity in the Brownian limit, and  a rescaled velocity distribution $F(z)$.
At first order we obtain
 \begin{equation}\label{eq:firstepR}
 \fl \int_0^\infty dy\; \int_0^{L/2} dx\; 
xy^2((1+\alpha_{+})\phi(x\Omega^*+y)
-(1+\alpha_{-} )\phi(x\Omega^*-y))=0,
 \end{equation} 
which can be solved numerically to obtain $\Omega^*$. At 
second order we obtain a 
Fokker-Planck equation for the rotor angular velocity distribution that gives, in the original variables
\begin{equation}
f(\Omega)\propto e^{-I(\Omega-\Omega^*)^2/2T_g}.
\end{equation}
where  $T_g/m$ is given by
\begin{equation}\label{eq:temprotor} 
\fl \frac{T_g}{m}=\frac{\int_0^{L/2} dx\;\int_0^\infty dy \,
x^2\,y^3((1+\alpha_{-})^2\phi(\Omega^*x\!-\!y)\!+\!(1+\alpha_{+}
)^2\phi(\Omega^*x\!+\!y)) }
{4\int_0^{L/2} dx\;\int_0^\infty dy\,
x^2\, y((1+\alpha_{-})\phi(\Omega^*x\!-\!y)+(1+\alpha_{+})\phi(\Omega^*x\!+\!y))}
\end{equation} 
For a rotor with $\alpha_{+}=1,\alpha_{-}=0$ in a Gaussian bath, equation (\ref{eq:MB}), we find
$T_g/T_B=0.72960$ which is close to the value for the corresponding asymmetric piston ($T_g/T_B=0.73460$).

\section{Torque-based approach}

The instantaneous impulse exerted on the rotor of angular velocity $\Omega$ by a
collision with 
a bath particle of velocity $v$ on a face with $\alpha_+$  is

\begin{eqnarray}
I_{-} &= I\Omega'- I\Omega\nonumber\\
&=-\frac{Imx}{I+mx^2}(I+\alpha_{-})(\Omega x-v), \;\;\; v<\Omega x
\end{eqnarray}
Similarly for a collision on a face with $\alpha_-$ 

\begin{equation}
I_{+} = \frac{Imx}{I+mx^2}(I+\alpha_{+})(v-\Omega x), \;\;\; \Omega x < v
\end{equation}
Assuming that successive collisions between bath particles and the granular motor are uncorrelated, 
averaging on different collisions at a given  $\Omega$ consists of 
integrating the impulse times the rate that the specified face collides
with a particle moving with a velocity $v$:
Evaluation of the torque requires an
integral over $x$, in addition to the bath particle velocity distribution, i.e. 
\begin{eqnarray}
\Gamma_{-}(\Omega)&=2\rho\int_0^{L/2}dx\;\int_{-\infty}^{\Omega x}dv\;
I_{-}(\Omega,v)(\Omega x-v)\phi(v)\\
\Gamma_{+}(\Omega)&=2\rho\int_0^{L/2}dx\;\int_{\Omega x}^{\infty}dv\;
I_{+}(\Omega,v)(v-\Omega x)\phi(v)
\end{eqnarray}

We suppose that, in addition to the fluctuating force resulting from
collisions with the bath particles, the rotor is subject to a
constant external torque, $\Gamma_{\rm ext}$. The average of the net torque
performed over all possible angular velocities of the rotor is equal
to zero in the stationary state. Since the successive
collisions are assumed uncorrelated, this condition can be expressed as,
\begin{equation}\label{eq:10} 
<\Gamma_{\rm net}(\Omega)>= \int_{-\infty}^{\infty} d\Omega f(\Omega)  \Gamma(\Omega) + \Gamma_{\rm ext} = 0
\end{equation}
where $\Gamma(\Omega)=\Gamma_{-}(\Omega)+\Gamma_{+}(\Omega)$.
It is convenient to rewrite this as
\begin{equation}
 <\Gamma_{\rm net}(\Omega)>=\Gamma(<\Omega>)+\Gamma_f+\Gamma_{\rm ext}=0
\end{equation} 
where $\Gamma(<\Omega>)$ is the mean torque when the rotor has constant
angular velocity, $<\Omega>$ and $\Gamma_f$ is the mean torque resulting
from the fluctuations of the angular velocity around the mean value, $<\Omega>$.
Now by expanding $\Gamma(\Omega)$ about $\Omega^*$, where $\Gamma(\Omega^*)=0$ we have that
\begin{eqnarray}\label{eq:pf}
\Gamma_f&=&<(\Omega-\Omega^*)\Gamma'(\Omega^*)+\frac{1}{2}(\Omega-\Omega^*)^2\Gamma''(\Omega^*)+\nonumber\\
&&+\frac{1}{6}(\Omega-\Omega^*)^3\Gamma'''(\Omega^*)+...>-\Gamma(<\Omega>)
\end{eqnarray}

So to lowest order the angular velocity is given by 
\begin{equation}\label{eq:expansion}
<\Omega> = \Omega^* -\frac{\frac{1}{2}\Gamma''(\Omega^*)<(\Omega-\Omega^*)^2>+\Gamma_{\rm ext}}{\Gamma'(\Omega^*)}
\end{equation}
Since $I<(\Omega-\Omega^*)^2>\approx T_g$, we have that in the Brownian limit,
$I\rightarrow\infty$, 

\begin{equation}\label{eq:vlim}
<\Omega> = \Omega^*-\Gamma_{\rm ext}/\Gamma'(\Omega^*)+O(1/I) 
\end{equation}

\begin{figure}
\begin{center}
\resizebox{8.0cm}{!}{\includegraphics{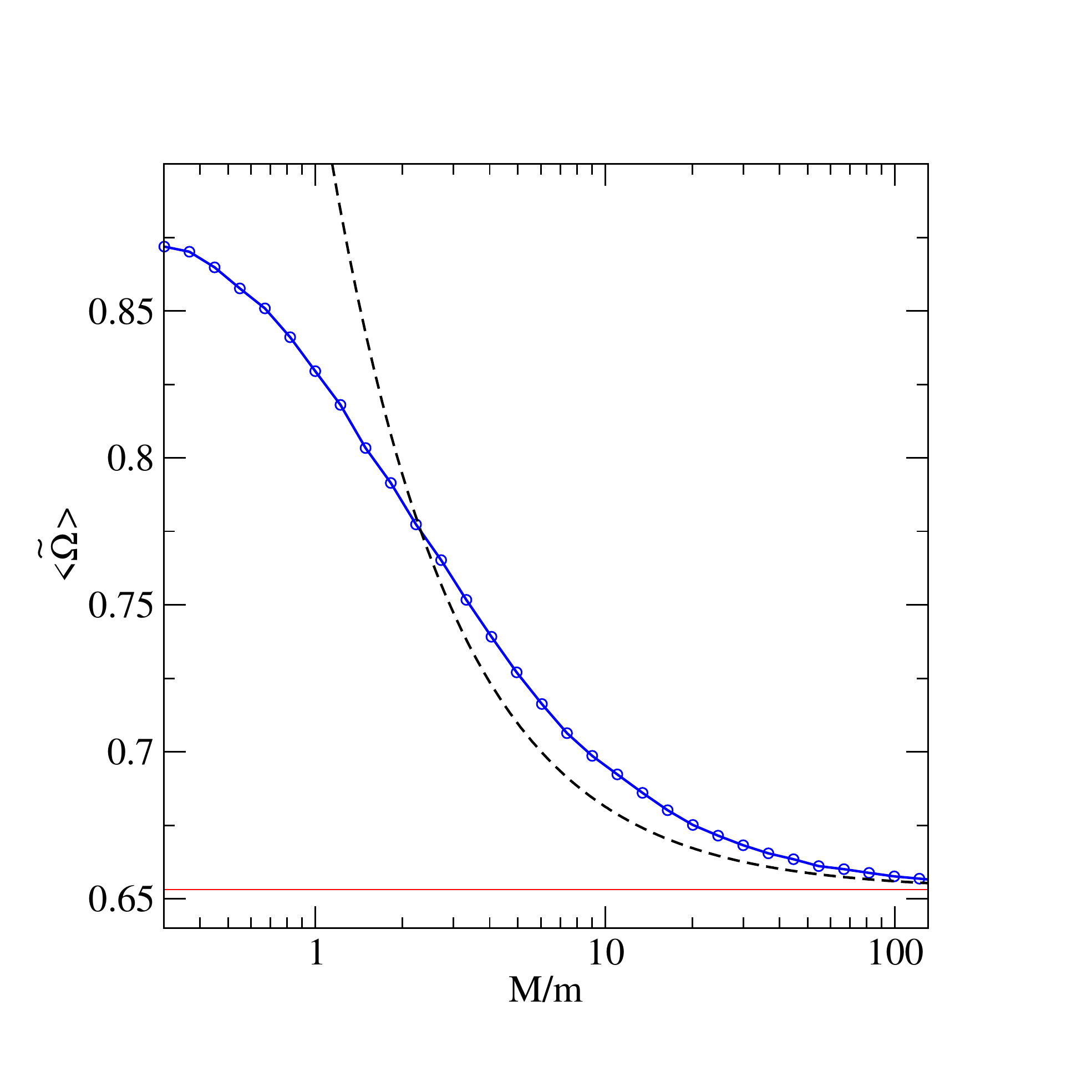}}
\end{center}
\caption{Dimensionless mean angular  velocity $\tilde{\Omega}$ as a function
  of the mass ratio $M/m$ for $\alpha_{+}=1$ and $\alpha_{-}=0$. The joined circles show the simulation results, the solid horizontal line
is the exact result in the Brownian limit (the solution of $\Gamma(\Omega)=0$ with $\Gamma(\Omega)$ given by equation (\ref{eq:gammagauss})) and the dashed line shows the first order
correction, equation (\ref{eq:omega1}).}
\label{fig:velo_rotor}
\end{figure}
Making the substitutions $y=\Omega x-v$ and $y = v-\Omega x$, we find the following expression for the total torque in the Brownian limit, $mL^2/I\rightarrow 0$:
\begin{equation}\label{eq:rotor}
 \fl\Gamma(\Omega)=2m\rho \int_0^{L/2}dx\int_{0}^{\infty}dy\;xy^2((1+\alpha_{+})\phi(x\Omega+y)-(1+\alpha_{-})\phi(x\Omega-y))
\end{equation}
We obtain the mean angular velocity, $\Omega^*$, 
by setting the torque equal to zero: $\Gamma(\Omega^*)=0$. 

The result, as well as simulations for finite ratios are shown in figure
\ref{fig:velo_rotor}. 
Note that, unlike the piston \cite{PhysRevE.82.011135}, we do not observe a maximum as the moment of inertia decreases. 

We can also calculate  corrections to the
Brownian limit by using equation (\ref{eq:expansion})
\begin{equation}\label{eq:omega1}
<\Omega>\simeq
\Omega^*-\frac{1}{2}\frac{\Gamma''(\Omega^*)}{\Gamma'(\Omega^*)}\frac{T_g}{I}
\end{equation}
$T_g$, $\Gamma'(\Omega^*)$ and $\Gamma''(\Omega^*)$ are functions of $\alpha_+$
and $\alpha_-$. Figure \ref{fig:velo_rotor} shows that, for $\alpha_-=0, \alpha_+=1$,
the first order correction provides a reasonably accurate
description for $M/m>5$, but for smaller values it diverges in contrast to 
the exact result that approaches a finite value.

\begin{figure}
\begin{center}
\resizebox{8.0cm}{!}{\includegraphics{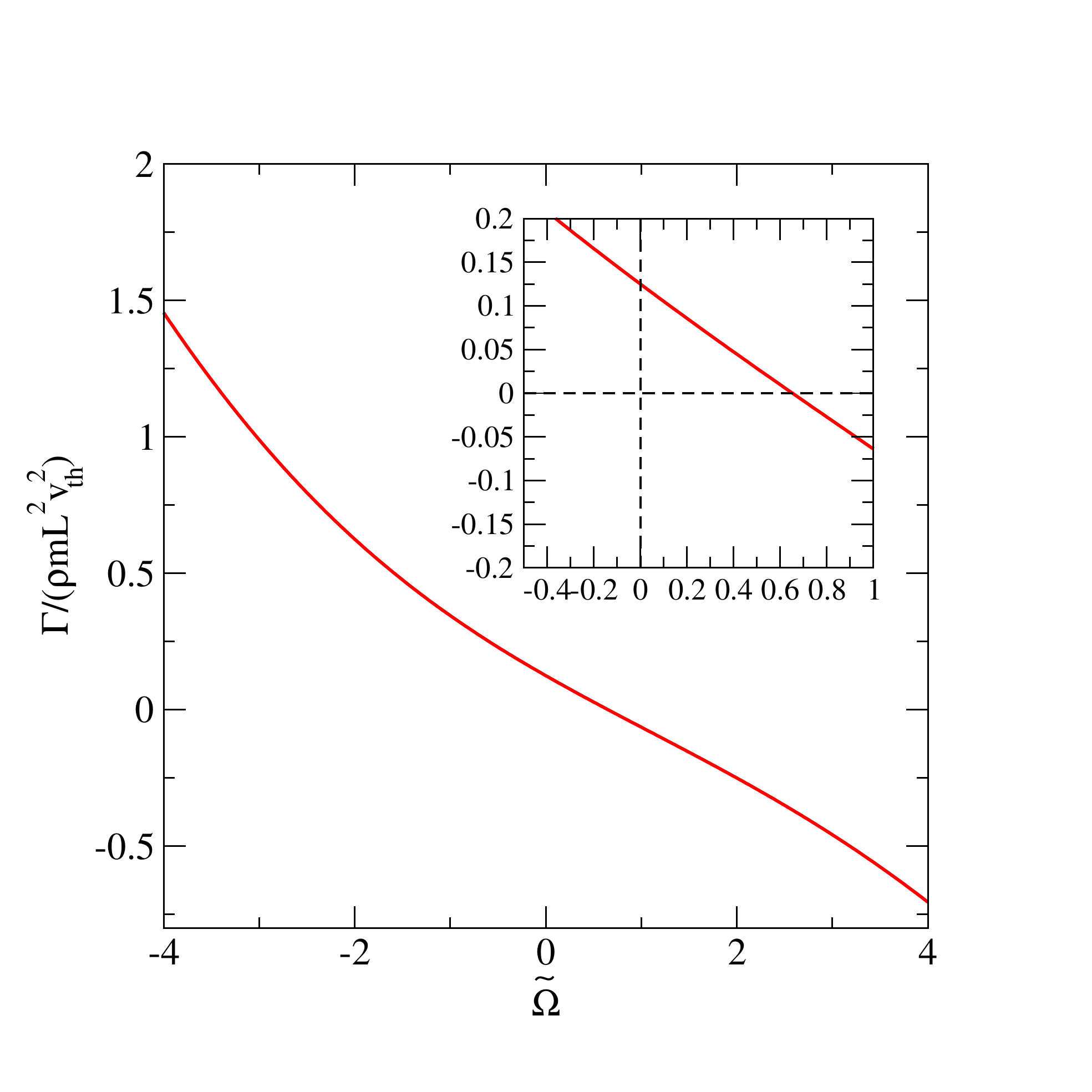}}
\end{center}
\caption{Dimensionless torque as a function
  of the mean angular velocity $\tilde{\Omega }$ for $\alpha_{+}=1$ and
$\alpha_{-}=0$. The
  torque is positive for $\tilde{\Omega}=0$ and vanishes for a positive value of
  $\tilde{\Omega} $, which corresponds to the solution of $\Gamma(\Omega)=0$ with $\Gamma(\Omega)$ given by equation (\ref{eq:gammagauss}). At
large
  angular velocity, the torque is opposite in sign to the angular velocity.}
\label{fig:torque}
\end{figure}

For an arbitrary symmetric bath distribution, the torque on a stationary rotor
in the Brownian limit is 
\begin{equation}\label{eq:torque0}
 \Gamma(0)=\frac{1}{8}m\rho L^2(\alpha_{+}-\alpha_{-})<v^2>
\end{equation}
In an experimental situation, this  must exceed the solid friction for the motor
to function. 

For small variations of the angular velocity, the torque exerced by the
particle bath  is proportional to the deviations from the steady
state $\Gamma(\Omega)=(\Omega^*-\Omega)\Gamma'(\Omega^*)$ where 
$\Gamma'(\Omega^*)$ is always a nonzero negative constant.

At large angular velocities, one obtains that 
\begin{equation}
 \Gamma(\Omega)\simeq \mp \frac{m(1+\alpha_\mp)\rho L^4}{32}\Omega^2, \;\;
\Omega \rightarrow \pm \infty
\end{equation}
The damping is quadratic at large values of the velocity (the piston has a similar behavior).

For a Gaussian distribution in the Brownian limit, we can obtain an analytical
expression for the torque:

\begin{eqnarray}\label{eq:gammagauss}
\fl \frac{\Gamma}{m\rho L^2 v_{th}^2}&=
\frac{\alpha_{+}-\alpha_{-}}{64}(\ot^2+8)-(2+\alpha_{+}+\alpha_{-})\nonumber\\
&\left[\frac{1}{64}(\ot^2+8-\frac{16}{\ot^2})erf(\frac{\ot}{2\sqrt{2}})
+\frac{1}{32}\sqrt{\frac{2}{\pi}}(\ot+\frac{4}{\ot})\exp(-\ot^2/8)\right]
\end{eqnarray}
where $\ot=\Omega L/v_{th}$ (see figure~\ref{fig:torque}). Expanding in powers of $\tilde{\Omega}$ to obtain

\begin{equation}
 \frac{\Gamma}{m\rho L^2 v_{th}^2}=\frac{1}{8}(\alpha_{+}-\alpha_{-})-\frac{1}{12}\sqrt{\frac{2}{\pi}}(2+\alpha_{+}+\alpha_{-})\ot+O(\ot^3)
\end{equation}
from which we can obtain an analytical estimate for the angular velocity in the steady state
\begin{equation}\label{eq:rotorapprox}
 \ot^*=\frac{3}{2}\sqrt{\frac{\pi}{2}}\frac{\alpha_{+}-\alpha_{-}}{2+\alpha_{+}+\alpha_{-}}
\end{equation}
We note that this  differs from the result for the asymmetric piston \cite{PhysRevE.82.011135} by a constant factor.
 This result also shows
that, for given coefficients of restitution and bath properties, the rotational velocity is inversely proportional to the length of the rotor.

\subsection{Dependence of the mean angular velocity on the coefficients of restitution}

\begin{figure}[t]
\begin{center}
\resizebox{8.0cm}{!}{\includegraphics{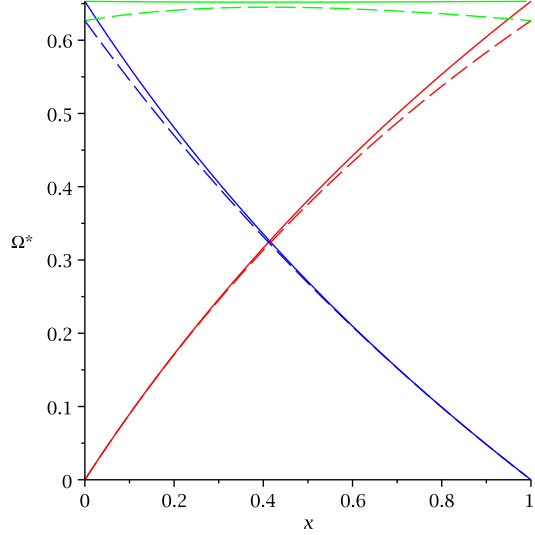}}
\end{center}
\caption{$\Omega^*(\alpha_{+}=x,\alpha_{-}=0)$ and $\Omega^*(\alpha_{+}=1,\alpha_{-}=x)$ and their sum as a
function of
$x$. The solid lines show the solution of $\Gamma(\Omega^*)=0$ with $\Gamma(\Omega)$ given by equation
(\ref{eq:gammagauss}) and the dashed lines show the approximate solution, equation
(\ref{eq:rotorapprox}).}
\label{fig:cor2}
\end{figure}

We examine the mean angular velocity in the Brownian limit as a function of the
coefficients of
restitution. It is easy to show starting from $\Gamma(\Omega^*)=0$ with $\Gamma(\Omega)$ given by
equation (\ref{eq:rotor}) 
that $\Omega^*(\alpha_{+}=x,\alpha_{-}=y)=-\Omega^*(\alpha_{+}=y,\alpha_{-}=x)$, 
regardless of the bath distribution. 
Of course this result is expected from simple symmetry considerations. 
In figure \ref{fig:cor2} we show $\Omega^*(\alpha_{+}=x,\alpha_{-}=0)$ and $\Omega^*(\alpha_{+}=1,\alpha_{-}=x)$
as a function of
$x$ in the case of a Gaussian bath distribution. The former increases as $x$
increases, while the latter decreases. 
They are equal for a particular value of $x$ and from equation (\ref{eq:gammagauss}) we may show
that this occurs for
$x=\sqrt{2}-1$. These results illustrate that the mean angular velocity is not simply
proportional to the difference 
in coefficients of restitution. Specifically, we can construct two rotors. The first is composed of an
elastic material ($\alpha_{+}=1$), and another material with a
coefficient of restitution of $x=\sqrt{2}-1$ or a difference of $2-\sqrt{2}=0.5858$. This has the 
same drift velocity as a second rotor composed of a completely inelastic material, ($\alpha_{-}=0$), and one with $\alpha_{-}=x$ or a difference of 0.4142.

These functions are very nearly symmetric i.e., to a good approximation 
they satisfy
\begin{equation}
\Omega^*(\alpha_{+}=x,\alpha_{-}=0)+\Omega^*(\alpha_{+}=1,\alpha_{-}=x)=\Omega^*(1,0) 
\end{equation}

Figure \ref{fig:cor2} also shows that the exact solutions are well-approximated by
the second order estimate equation (\ref{eq:rotorapprox}).

\subsection{Power and efficiency}

A defining characteristic of granular motors is their ability to extract energy from the bath in the form
of mechanical work. We now proceed to the calculation of the mechanical work, as well as 
the dissipated power due to the inelastic collisions between the granular motor and the bath 
particles. The ratio of these two quantities defines the efficiency of the motor.

If the external torque in equation (\ref{eq:10}) is non-zero, work is being done by, or on, the 
rotor. The power is given by
\begin{equation}\label{eq:powerdef}
\dot{W}=\Gamma_{\rm ext}<\Omega>
\end{equation}
In the Brownian limit we can estimate the average drift velocity using equation (\ref{eq:vlim}).
Alternatively, we can take $<\Omega>$ as the independent variable: 
\begin{equation}\label{eq:power}
\dot{W}=\Gamma(<\Omega>)<\Omega>
\end{equation}
with $\Gamma(\Omega)$ given by equation (\ref{eq:rotor}). Of course, this approach is only valid in the Brownian limit.
In general $\dot{W}(\Omega)$ has a parabolic form and is maximum for  $0<\Omega<\Omega^*$. When the angular 
velocity is greater than $\Omega^*$ the power
is negative, implying that it is necessary to drive the rotor externally in order to maintain the motion.

\begin{figure}[t]
\begin{center}
\resizebox{8cm}{!}{\includegraphics{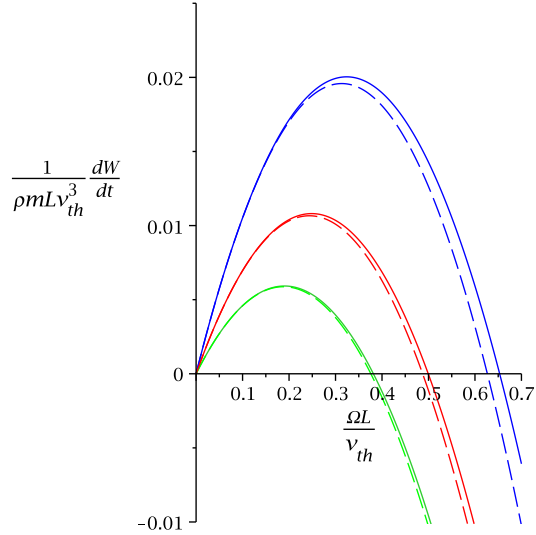}}
\end{center}
\caption{Power as a function of the rotor angular velocity for a system with
  $\alpha_{-}=0$ in the Brownian limit. The solid and dashed lines show the exact (equation (\ref{eq:gammagauss}) in equation (\ref{eq:power})) and
  approximate (equation (\ref{eq:power2})) values, respectively.  $\alpha_{+}=1,0.7,0.5$ in order of
  decreasing amplitude}
\label{fig:power}
\end{figure}

At second order

\begin{equation}\label{eq:power2}
 \frac{\dot{W}}{m\rho Lv_{th}^3}=\frac{1}{8}(\alpha_{+}-\alpha_{-}) \ot-\frac{1}{12}(2+\alpha_{+}+\alpha_{-})\sqrt{\frac{2}{\pi}}\ot^2
\end{equation}
Figure~\ref{fig:power} shows that this provides a good approximation.
The maximum power occurs at $\Omega_{mp}=\Omega^*/2$ and is given by

\begin{equation}
 \frac{\dot{W}_{\rm max}}{m\rho Lv_{th}^3}=\frac{3\sqrt{2\pi}}{128}\frac{(\alpha_{+}-\alpha_{-})^2}{2+\alpha_{+}+\alpha_{-}}
\end{equation}

From equation~(\ref{eq:energy}), we find that  the rate of energy dissipation resulting from collisions
 between the bath particles and the rotor is
\begin{equation}
 \fl\dot{E}_{tot}(\Omega)=-m\rho\int_0^{L/2}dx\int_0^{\infty}dy\;y^3[(1-\alpha_{-}^2)\phi(\Omega x - y)+(1-\alpha_{+}^2)\phi(\Omega x + y)]
\end{equation}

\begin{figure}[t]
\begin{center}
\resizebox{8cm}{!}{\includegraphics{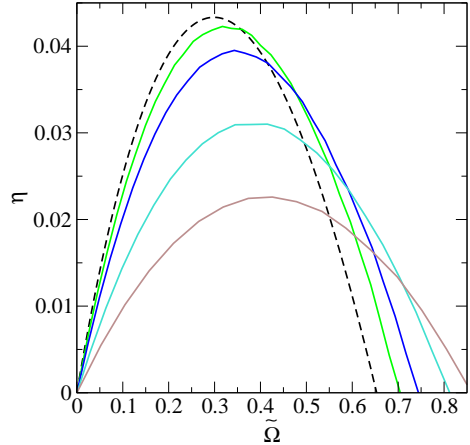}}
\end{center}
\caption{The efficiency as a function of the dimensionless mean drift angular velocity for a system with
  $\alpha_{-}=0,\alpha_{+}=1$. The solid curves show simulation
  results for $M/m=1,2,5,10$ (bottom to top), while the dashed line
  shows the theoretical prediction in the Brownian limit. The bath particle velocity distribution is given by equation (\ref{eq:MB}).
}
\label{fig:effrotor2}
\end{figure}

The efficiency, or the fraction of the dissipated energy that is converted into work, is:
\begin{equation}
\eta(\alpha_{+},\alpha_{-},\Omega) = \frac{\dot{W}}{|\dot{E}_{tot}|}
\end{equation}
For an arbitrary bath distribution in the Brownian limit we find that
\begin{equation}\label{eq:effrotor}
\fl\frac{\eta(\Omega,\alpha_{+},\alpha_{-})}{2\Omega}=
\frac{\int_0^{L/2}dx\int_{0}^{\infty}dy\;xy^2((1+\alpha_{+})\phi(x\Omega+y)-(1+\alpha_{-})\phi(x\Omega-y))}{\int_0^{L/2}dx\int_{0}^{\infty}dy\;y^3((1-\alpha_{+}^2)\phi(x\Omega+y)+(1-\alpha_{-}^2)\phi(x\Omega-y))}
\end{equation}
For a Gaussian bath distribution this has a maximum value of 4.3335\% for
$\alpha_{+}=1,\alpha_{-}=0,\tilde{\Omega} =0.29864$. 
Figure~\ref{fig:effrotor2} shows the efficiency as a function of the dimensionless mean angular velocity obtained by simulation for several mass ratios,
 $M/m$ as well as the
theoretical result in the Brownian limit, equation (\ref{eq:effrotor}).

This mechanical analysis indicates that heterogeneous granular particles are able to produce 
significant noise rectification. 
Provided that the solid friction
about the axis is not too large, the chiral rotor should be a good candidate
for an experimental realization of a Brownian granular motor.

\section{Generalization to an n-vaned rotor}

If one retains the assumption that the bath remains homogeneous and is not influenced by the presence 
of the rotor, it is easy to generalize the above analysis to a rotor composed of $n$ vanes. 
In the Boltzmann-Lorentz equation~(\ref{eq:berot}), the factor of $1/2\rho$ on the left hand side is replaced by $1/n\rho$. 
In the steady state the entire left hand side is equal to zero, so the steady state angular velocity is unchanged. 
In the expression for the torque, the factor of $2$ in equation (\ref{eq:rotor}) is replaced by $n$. But since this factor does not
affect the solution of $\Gamma(\Omega^*)=0$, the number of vanes has no effect on $\Omega^*$. We note, however, that 
the assumption of bath homogeneity certainly deteriorates as the number of vanes increases.

\section{A numerical experiment}

The analysis presented in the preceeding sections assumes that
the bath is perfectly homogeneous. To investigate effect of relaxing this assumption, we now
consider a realistic simulation of a quasi two-dimensional rotor.
Burdeau {\it et al}.\cite{PhysRevE.79.061306} developed a
model that accurately reproduces an experimental study of a vibrated
bilayer. The first layer consists of densely packed heavy granular particles while the second layer is composed of light particles
with an intermediate coverage. Both experimentally and in the simulation, one observes
a horizontal velocity distribution function of the
light particles that is very close to a Gaussian \cite{PhysRevE.79.061306}.

\begin{figure}[t]
\begin{center}
\resizebox{8.0cm}{!}{\includegraphics{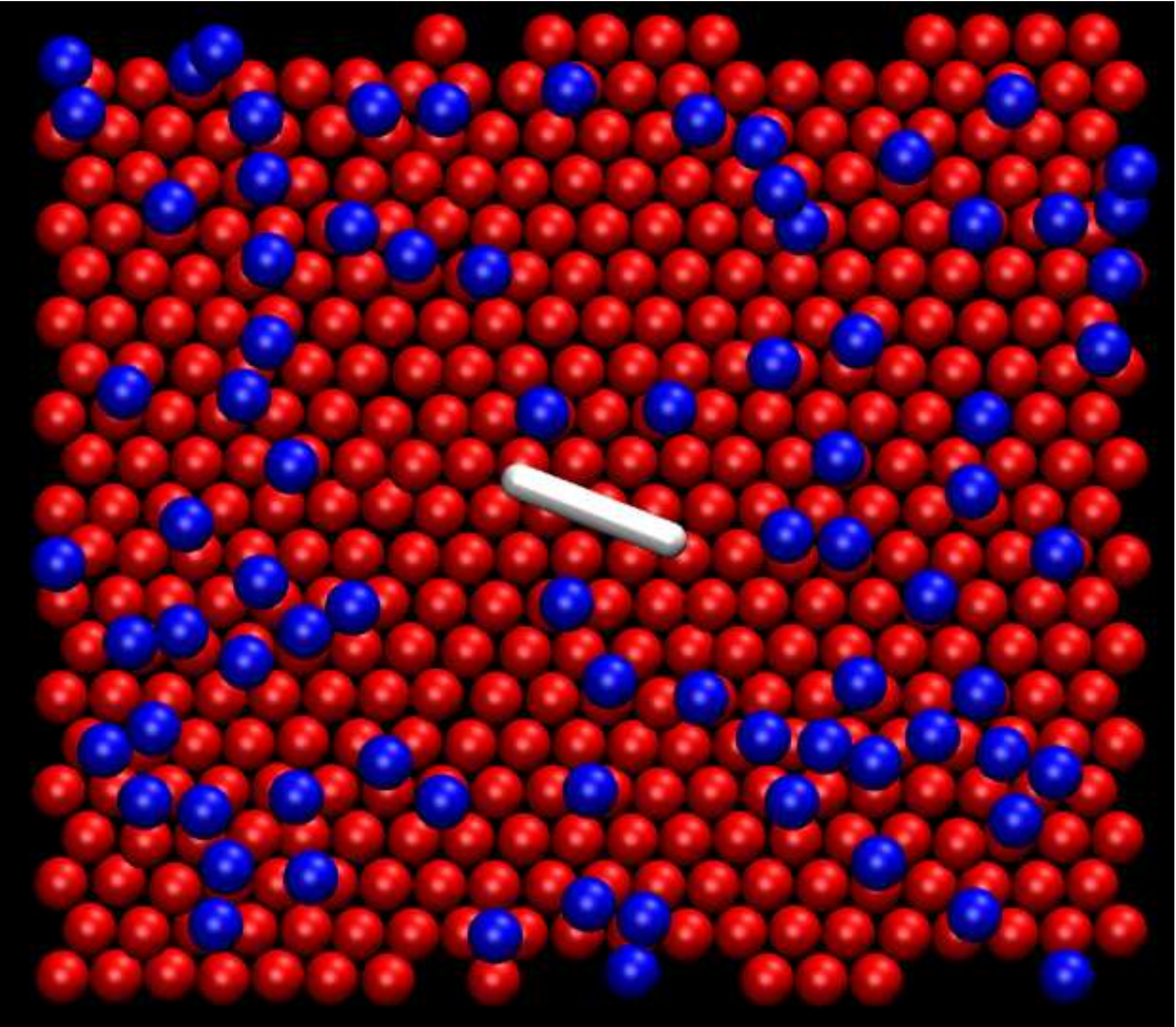}}
\end{center}
\caption{Top view of the simulation cell of size $20\times 20$ sphere
diameters: The spherocylinder 
has a radius $R=2/3r$ ($r$ is the radius of the spheres) and a total length $L+2R$ where $L=11R$.}
\label{fig:snap_sphero}
\end{figure}

\begin{figure}[th]
  \begin{center}

\includegraphics{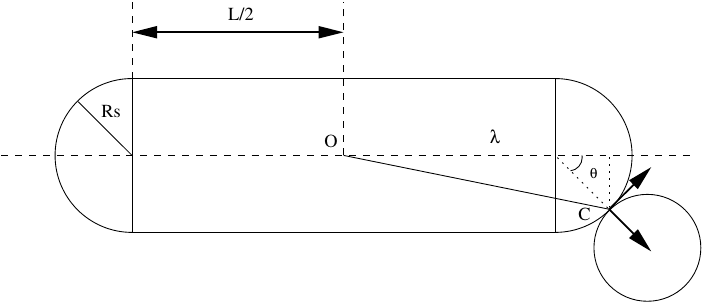}%
\end{center}
\caption{Geometry of  the spherocylinder and a sphere  during a collision}
\label{fig:geometry}
\end{figure}

To this vibrating bilayer system, we have added here a three dimensional spherocylinder
constrained to rotate around a vertical axis: See figure \ref{fig:snap_sphero}.  The spherocylinder cannot move
horizontally or vertically: the rotation around the Oz axis is the sole degree
of freedom.  Figure \ref{fig:geometry} is a projection of the spherocylinder together with the geometrical parameters.

The simulation uses a Discrete Element Method where collisions between particles, collisions between spherocylinder and 
particles as well as collisions with the vibrating plate are all inelastic. In  addition, the particles are  subject to a constant acceleration due
to the (vertical) gravitational  field.  The visco-elastic forces  are
modeled  by the  spring-dashpot  model \cite{cundallstrack79}. 

The systems consists of  $N_1$   spheres placed in the first layer  and $N_2$   spheres in  the
second layer. 
By choosing  the system parameters  as close as
possible to the original experimental system\cite{Baxter2003},  no mixing is observed between layers and the
horizontal velocity distribution of the second layer is very close to a Gaussian.
 The  force  along the line joining  the
two  centers  that dissipation into  account  is a
damped harmonic oscillator defined as:
\begin{equation}
F_{n}=-k_{n}\xi-\gamma_{n}\dot{\xi}
\end{equation}
where $\xi$ is the overlap between two particles, $k_{n}$   is  related to  the  stiffness  of the  material,  and
$\gamma_{n}$ to the dissipation.  This force model  allows one to very
easily  tune  certain quantities  in   the simulation, namely  the
normal  coefficient of restitution $e_{n}$ (which  is then constant for all
collisions at all velocities in our system). Relevant values of $e_{n}$
and $t_{n}$   determine   the  values of  $k_{n}$  and   $\gamma_{n}$,
independently of the velocities.  The collision duration,
$t_{n}$,  provides a microscopic  characteristic time. The
simulation time step is taken as $\Delta t = 10^{-5}s$ and the mean
duration of a collision as $t_{n} = 10^{-3} s$. 
A frictional force  between spherical particles is added by taking the
tangential
component of the force as $F_{t}= - \rm{min}(|k_{t}\zeta|,|\mu F_{n}|)$
where $k_{t}$ is  related to the tangential  elasticity and $\zeta$ is
the    tangential  displacement   when    the   contact   was  first
established. We   used  a  ratio $k_{t}/k_{n}=2/7$ with $\mu=0.25$.
In order that the spherocylinder only undergoes collisions  with particles of the second layer, we choose the
vertical height of its center of gravity as $h=3r+R$. 

We first performed simulations for a homogeneous spherocylinder, where no mean angular velocity appears.
If the linear dimension of of the simulation cell is at least twice the length of the
spherocylinder,  the boundary conditions do not influence  the kinetic properties of the latter.
We then considered a chiral spherocylinder made of two materials of coefficient of restitution $\alpha_+=0.8$ and
$\alpha_-=0.2$. 

\begin{figure}[th]
\begin{center}
\resizebox{8cm}{!}{\includegraphics{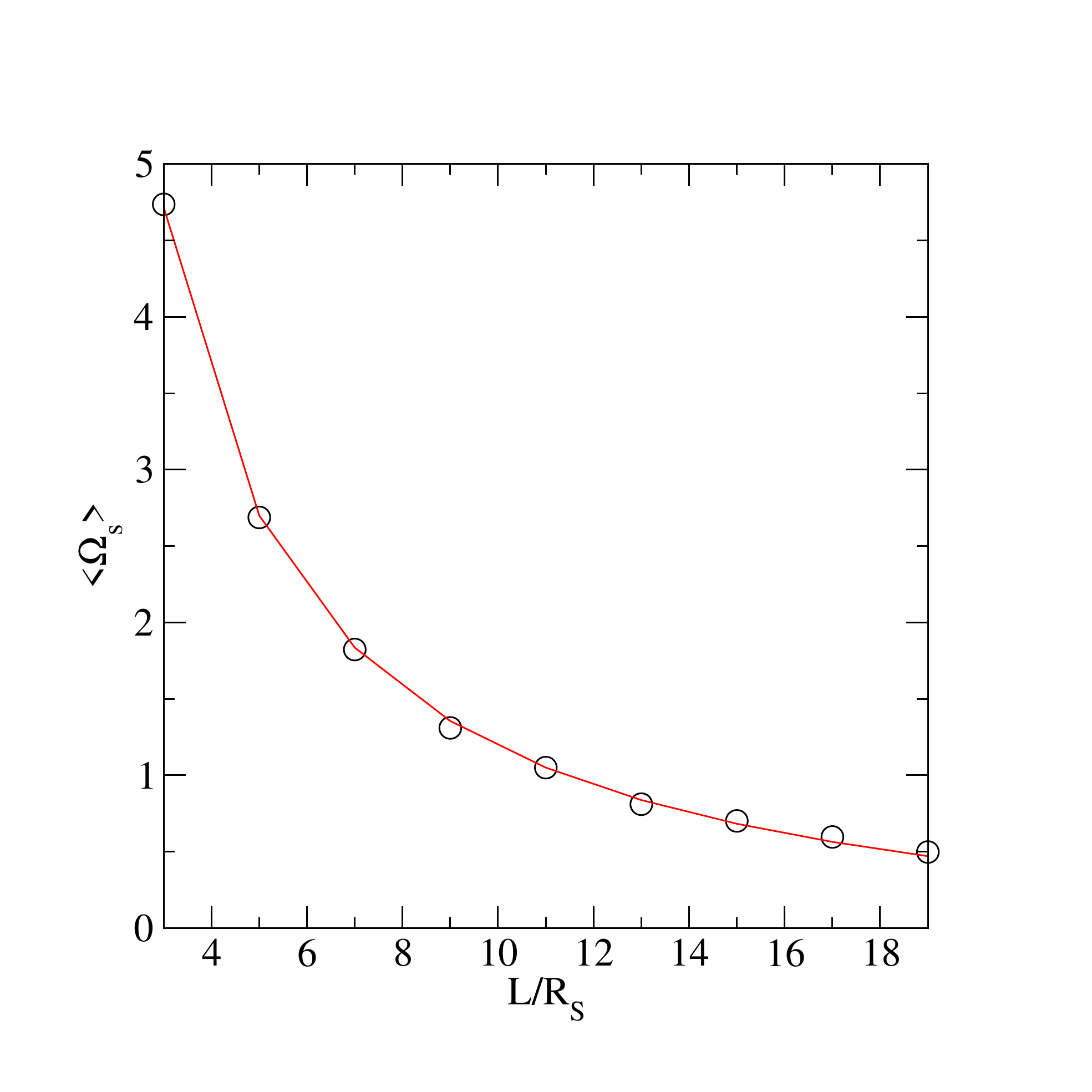}}
\end{center}
\caption{Mean angular velocity versus the length of the spherocylinder with $\alpha_+=1$, $\alpha_-=0.25$, 
and  spherocylinder mass $M_S=40m$.}
\label{fig:omega}
\end{figure}

\begin{figure}[th]
\begin{center}
\resizebox{8cm}{!}{\includegraphics{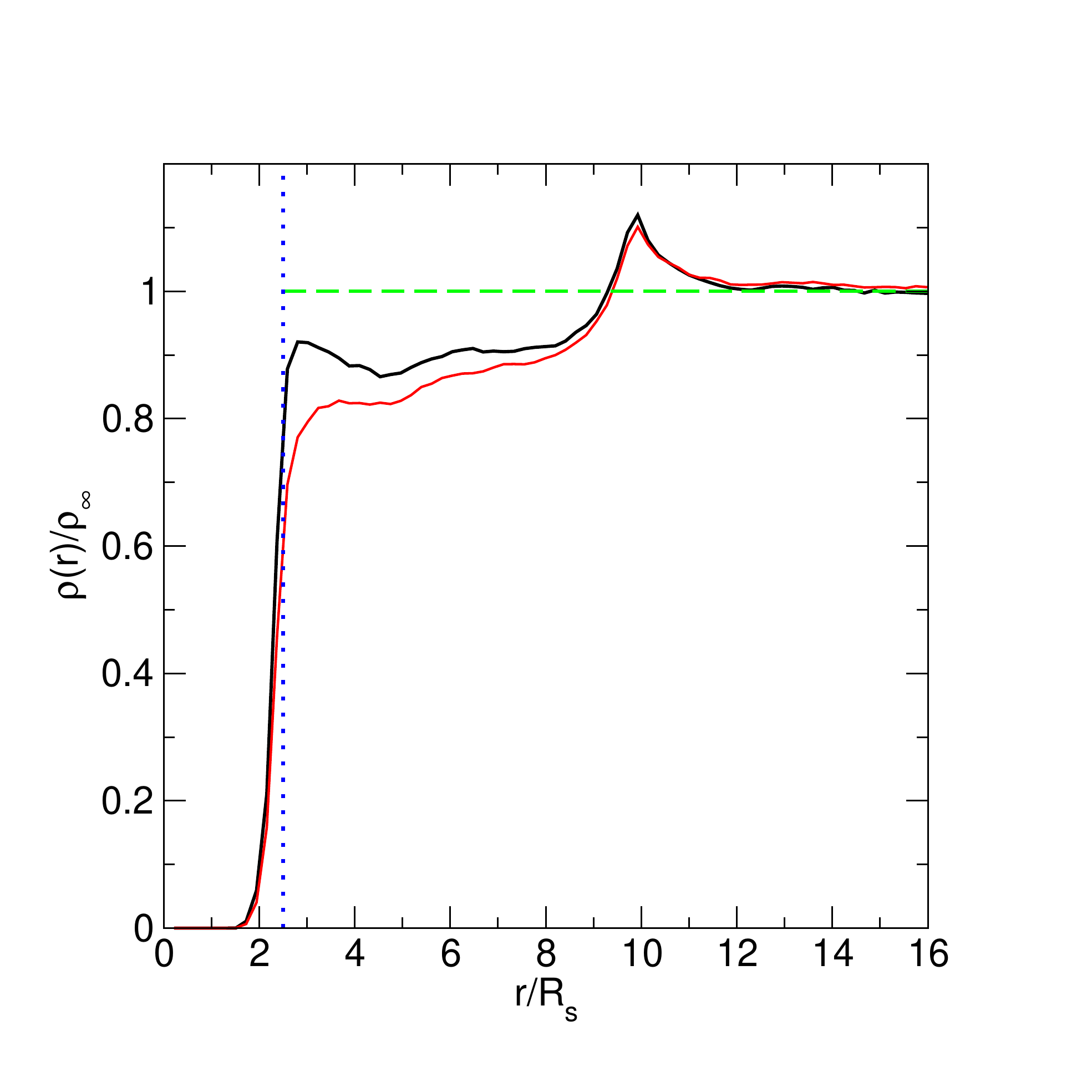}}
\end{center}
\caption{Dimensionless density of the bath particles (second layer) versus  the dimensionless distance to the
spherocylinder center  $r/R_s$, with a bath density corresponding to a two-dimensional coverage of 0.18. 
The bath particle radius  is $R=1.5R_s$. The black and red curves correspond to homogeneous ($\alpha=0.2$) 
and heterogeneous ($\alpha_+=0.8, \alpha_-=0.2$)
spherocylinders, respectively. The total length of the spherocylinder is $L+R_s=16R_s$, 
The dashed line represents the Boltzmann-Lorentz assumption of a perfect homogeneous bath.}
\label{fig:density}
\end{figure}

The motor effect is clearly evidenced in this numerical experiment: See figure \ref{fig:omega}. The Boltzmann-Lorentz approach can be
extended to objects of finite area (see Ref.~\cite{GTV05}). In the limit of large $L$ compared to the bath particle radius and the radius of the cap, 
the mean angular velocity varies as $1/L$ as shown in figure \ref{fig:omega}. We also monitored 
the density of the bath particles in the region of the spherocylinder. Figure \ref{fig:density} shows the near exclusion of particle centers 
in the range $0<r<R+R_s$ due to the steric  effect (the fact that the density is not strictly zero is due to out-of-plane collisions resulting from the
three dimensional nature of the system). The vertical dotted line corresponds to $r=R+R_s$. 
One observes a small depletion in the region swept out by the rotating spherocylinder followed by an increase to a maximum 
at a distance close to $(L+R_s)/2+R=9.5R_s$ i.e., a distance 
where bath particles collide with the ends of the spherocylinder. Beyond this distance the density rapidly approached the bulk value. 
These effects are similar for homogeneous and heterogeneous spherocylinders. 
For comparison the straight dashed line corresponds to the assumption of a homogeneous bath used in the Boltzmann-Lorentz description.  

\section{Conclusion}

We have applied kinetic theory and realistic numerical simulation to study the properties of
a chiral  heterogeneous granular motor. As with the previously studied piston, a strong ratchet effect may be produced under certain conditions. 
We also developed a mechanical analysis that provides an exact description in the Brownian limit 
where the motor is much more massive  than a bath particle. 

The mechanical approach may also be used to assess the robustness of the ratchet effect. 
In an experimental situation, a dry friction force between the rotor axle and its bearing is present. 
If the force or torque acting on a stationary motor exceeds the static friction, the motor effect
should be present. In their experiment 
Eshuis et al.\cite{PhysRevLett.104.248001} observed that  
there is a net rotation of the asymmetric vanes that is only non-zero if the bath particles
are sufficiently agitated. This is consistent with our analysis that gives an expression
for the torque acting on a stationary rotor, equation (\ref{eq:torque0}). If this is larger than 
the static friction, the rotor starts to turn. If the rotor is immobile the granular temperature
of the bath particles can be increased by increasing the vibration amplitude or frequency until the
threshold is reached. 

Eshuis et al.\cite{PhysRevLett.104.248001} also observed a convective motion that accompanies the motor rotation. 
This is different from the effect observed in a system of vertically vibrated glass beads
where a toroidal convection roll with a vertical axis appears\cite{WHP01a}. The origin of this effect is the
inelastic  collisions between particles and side wall\cite{TV02}. In the experiment of Eshuis et al., however, 
convection accompanies the rotation of the vanes and is not the origin of the motor effect. 

A more quantitative analysis of the influence
of dry friction on the ratchet effect requires its incorporation in the kinetic description. 
We are currently pursuing this direction.

We thank  J. Piasecki for useful discussions.
\section*{References}



\end{document}